# Optical modulation of Gate-Induced Electron Trapping via Persistent Photoconductivity in SrTiO$_3$/AlO$_x$ Heterostructures

Peiwen Luo, Huizhong Zeng, Bin Peng, Wanli Zhang and Wenxu Zhang

*National key laboratory of electronic thin films and integrated devices, University of Electronic Science and Technology of China, Chengdu 611731, China*

The dynamic interplay between light and electric field control of charge states lies at the heart of developing multifunctional optoelectronic devices. While persistent photoconductivity (PPC) and gate-voltage ($V_G$)-induced electron trapping are well-known phenomena in oxide heterostructures, their mutual coupling remains poorly explored. Here, we report that the non-equilibrium state established by PPC can effectively modulate the efficacy of $V_G$-induced electron trapping in a SrTiO$_3$/AlO$_x$ heterostructure. The PPC, characterized by a slow relaxation ($\tau_l$ = 9 hours at 4 K) after sub-illumination, originates from the re-trapping of photoexcited carriers into deep-level states. In contrast, $V_G$-induced trapping, governed by shallow states, exhibits much faster dynamics ($\tau \sim 100$ s). Crucially, we discover that the strength of $V_G$-induced trapping is not constant but is dynamically modulated by the PPC relaxation process. The trapping amplitude is strongly amplified after illumination and recovers only after the deep-level states are substantially refilled, precisely following the PPC relaxation. Furthermore, the electron trapping effect diminishes with increasing temperature and vanishes near the ferroelastic phase transition of STO (~110 K), confirming that ferroelastic twin walls and associated oxygen vacancy clusters are the physical origin of the traps. Our findings reveal a novel optical gating mechanism for electron trapping, paving the way for designing non-volatile, optically programmable electronic devices.

## I. INTRODUCTION

Persistent Photoconductivity (PPC), a remarkable optoelectronic phenomenon observed in numerous semiconductors, is characterized by a long-lived enhancement of electrical conductivity following the cessation of light illumination[1]. This effect, which can persist from seconds to days, is conventionally attributed to the trapping of photo-generated carriers by metastable defects[2]. In metal oxides, oxygen vacancies (OVs) often serve as such defects, where substantial lattice relaxation under illumination[3] creates a high energy barrier against electron–hole recombination. This structural reconfiguration effectively "locking" electrons in the conduction band and resulting in a light-triggered non-volatile conductive state[4]. This inherent memory effect makes PPC of technological significance in neural devices. The phenomenon becomes profoundly more complex at the interfaces of strongly correlated oxide heterostructures like SrTiO$_3$ (STO), where intricate coupling among charge, spin, orbital, and lattice gives rise to a rich spectrum of electronic phenomena[5, 6]. At these interfaces, the coexistence of abundant defect states and strong built-in electric fields makes the interface highly responsive to light illumination, enabling giant PPC even at room temperature[7, 8]. Crucially, in such systems, PPC does not occur in isolation but is intrinsically coupled with other tuning parameters, particularly electrostatic gating[9]. A gate voltage ($V_G$) applied to an STO-based heterostructure directly modulates the interfacial potential and carrier density. More importantly, it alters the occupation energy and filling probability of defect states which act as trapping centers for mobile carriers[10]. This suggests a fundamental link between the persistent photocurrent and the gate-bias-modulated electron trapping, as both phenomena govern the 2DEG by modifying the filling state of the same interface trap spectrum. While PPC and gating effect in STO heterojunctions have been studied as independent phenomenon, their interactions as a coupled system remains poorly explored. Only recently, it is demonstrated that a PPC-induced low-resistance state can be electrically programmed by a gate voltage[11-13], highlighting the potential for optoelectronic control.

In this paper, we focuses on the interaction between PPC and $V_G$-induced electron trapping in the STO/AlO$_x$ heterostructure. Unlike the electronically reconstructed STO/LAO interface, the 2DEG in STO/AlO$_x$ originates from a redox reaction between TiO$_2$ and Al[14], which generates a high density of OVs at the interface. We demonstrate that the $V_G$-induced trapping arises primarily from shallow defect states, and can be significantly modulated by the non-equilibrium state established by PPC. Conversely, the additional carriers induced by gating also serve as a sensitive probe of the PPC process. By monitoring the trapping dynamics of those carriers, we unveil the gradual transition from deep- to shallow-level dominated trapping during the PPC recovery. Our findings provide an experimental foundation for designing novel, multifunctional photoelectric oxide devices.

## II. EXPERIMENTAL DETAILS

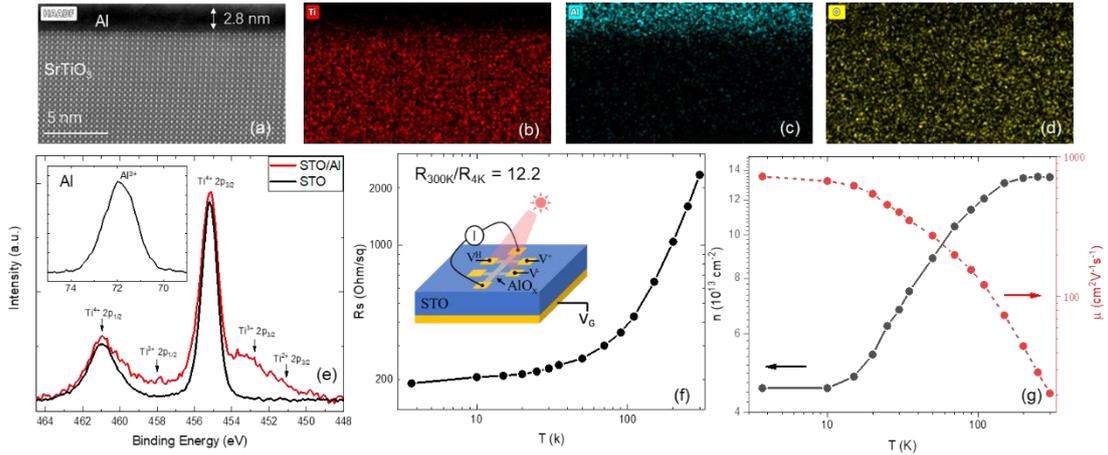

Fig. 1 (a) STEM image of a STO/AlO$_x$ heterostructure. (b-d) EDX maps for Ti, Al and O. (e) XPS of Ti 2p state for a bare STO before (black line) and after (red line) deposition of 3nm of Al. The inset is the spectrum for the Al 2p state on the STO/AlO$_x$ sample. (f) Temperature dependence of resistance of the STO/AlO$_x$ heterostructure. The inset is the sketch of the Hall experiment. (g) Temperature dependence of carrier density and mobility of the STO/AlO$_x$ heterostructure.

The sample was fabricated by depositing an approximately 3 nm thick aluminum layer onto a TiO$_2$-terminated STO (001) substrate via DC sputtering. The deposition was carried out at a substrate temperature of 350 °C, with a sputtering power of 20 W under an argon atmosphere at a pressure of 0.2 Pa. A representative cross-sectional high-angle annular dark-field (HAADF) scanning transmission electron microscopy (STEM) image of a resulting STO/AlO$_x$ heterostructure is presented in Fig. 1(a). The image reveals a continuous Al layer of uniform thickness on the STO substrate. Energy-dispersive X-ray spectroscopy (EDS) elemental maps for Ti, Al, and O are shown in Fig. 1(b-d). The redox reaction between Al and the TiO$_2$ layer extracts oxygen from the STO surface, reducing Ti and oxidizing Al. The Al and O elemental maps confirm complete oxidation of the Al layer, forming an amorphous AlO$_x$ layer[15].

X-ray photoelectron spectroscopy (XPS) is performed on both a bare STO substrate and the STO/AlO$_x$ heterostructure, focusing on the Ti 2p and Al 2p states (Fig. 1e). The spectrum collected from the bare STO (black line) corresponds to a Ti$^{4+}$ valence state, indicating its insulating character. After Al deposition, peaks associated with Ti$^{3+}$ and Ti$^{2+}$ valence states appear, confirming Ti reduction and oxygen vacancy formation[16, 17]. The Al 2p spectrum of the STO/AlO$_x$ heterostructure shows a single peak associated with Al$^{3+}$ valence state (inset of Fig. 1e), demonstrating that the Al layer has been totally oxidized by the air, which act as a protective layer to prevent the 2DEG from being oxidized.

Temperature-dependent sheet resistance ($Rs$) measurements from 4 K to 300 K are shown in Fig. 1(f). $Rs$ remains nearly constant below 50 K and increases significantly up to 300 K, indicating its metallic behavior. The residual resistance ratio (RRR = $Rs$(300 K)/$Rs$(4 K)), used to characterize metal purity, is as high as 12, suggesting minimal lattice distortion and low impurity concentration at the interface[18]. Charge density ($n_s$) and Hall mobility ($\mu$) were measured using a standard six-electrode Hall bar geometry (inset of Fig. 1f), with a channel length and width of 1.6 mm and 0.4 mm, respectively. The temperature dependence of $n_s$ and $\mu$ for the 2DEG is shown in Fig. 1g. As mentioned above, the 2DEG primarily originates from the interface redox reaction and OVs. Increasing temperature accelerates OV ionization, releasing additional electrons to the 2DEG and increasing $n_s$[19]. The saturation of $n_s$ around 150 K suggests complete OV ionization. Conversely, ionized oxygen vacancies act as charged scattering centers, reducing carrier mobility (red line in Fig. 1g). Furthermore, the temperature dependence of $n_s$ can be characterized by an activation energy $\varepsilon$ of ~5 meV using the thermal excitation model $n_s \propto \exp(-\varepsilon/k_B T)$,[19] where $k_B$ is the Boltzmann constant. This result lead us to infer that the relevant OVs form shallow defect states located near the conduction band minimum of STO[20]. As will be shown later, such shallow states are effectively modulated by a gate voltage, whereas the release of electrons from deep levels, such as those identified in related oxide heterostructures at approximately –1.2 eV and –2.1 eV below the conduction band[21], requires higher energy excitation. Although these deep states contribute negligibly to conduction under thermal equilibrium, they serve as the key reservoirs responsible for the persistent photoconductivity observed in this system, as demonstrated in the following sections.

## III. RESULTS AND DISCUSSION

### A. Electron trapping induced by gate voltage and light illumination

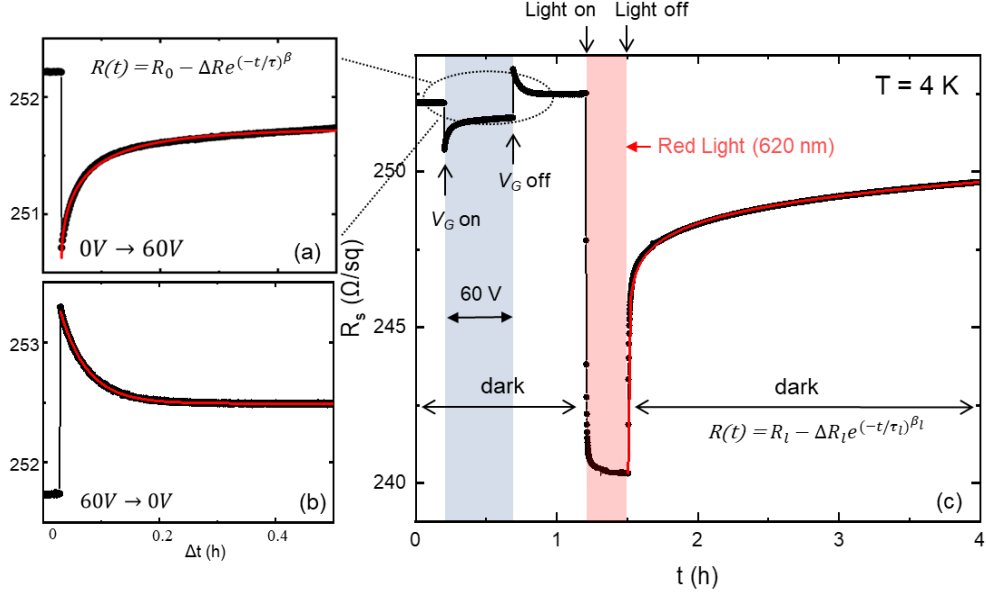

Fig. 2 (a) $R_S$ variation with time under positive and (b) negativize gate voltage step at 4 K, the red line is the fitting curve of the electron trapping and de-trapping process. The inset of Fig. (a) is the formular (Eq. 1) used for fitting ($\tau$ = 150 s, $\beta$ = 0.9). (c) light-induced variation of $R_S$ with time (black line), the red area represents the light illumination with a wavelength of 620 nm. The red line is the fitting curve of the PPC process with $\tau_l$ = 9 h and $\beta_l$ = 0.2.

To explore $V_G$-induced electron trapping in the 2DEG, we first applied a single voltage pulse sequence (0 V → 60 V → 0 V) to the STO/AlO$_x$ heterostructure at 4 K, with each voltage step lasting approximately 30 minutes. The corresponding variation of $R_S$ with $V_G$ is shown in Fig. 2(a) and (b). Due to electrostatic charging, applying a positive gate voltage step (+60 V) initially causes a sudden drop in $R_S$[22]. Surprisingly, $R_S$ then undergoes a gradual increase over several minutes. Similar $R_S$ relaxation has also been observed in other oxide heterostructures like STO/LAO[10, 22] and STO/GAO[23] attributed to the trapping of free carriers by trap centers formed by OV clusters and STO ferroelastic twin walls. Accordingly, the relaxation of $R_S$ can be described by the Kohlrausch-Williams-Watts (KWW) function[9]:

$$R(t) = R_0 - \Delta R e^{(-t/\tau)^\beta}, \qquad (1)$$

as shown by the red line in Fig. 2(a). Here, $\tau$ is the relaxation time constant and $\beta$ quantifies the stretching of the decay, which takes values between 0 and 1. A value of $\beta$ = 1 corresponds to a single-exponential relaxation, whereas smaller $\beta$ indicates a broader distribution of relaxation time. The fitted parameters yield $\tau$ = 150 s and $\beta$ = 0.9. The time constant $\tau$ is similar to reports in other oxide heterostructures[10, 23] but longer than in conventional semiconductor devices[24], indicating that the trapping effect in oxide heterostructures is a slow process due to the involvement of OVs. The large value of $\beta$ suggests that the trapping centers involved are energetically narrow and locate at a similar energy level within the gap. Similarly, applying a negative voltage step (60 V → 0 V) causes a sudden rise in $R_S$ followed by a gradual decrease (Fig. 2b), which is also well fitted by the Eq. (1) with with comparable parameters $\tau$ = 160 s and $\beta$ = 0.9, consistent with a reversible de-trapping process from OV-related defect states.

Fig. 2(c) shows the $R_S$ variation under the red light (620 nm) illumination. As expected, $R_S$ drops sharply under illumination. Since the photon energy (~2 eV) is less than the STO bandgap (3.2 eV), this drop primarily originates from the excitation of localized electrons from within the band gap[25]. Recent studies on GAO/STO heterostructures identified in-gap states at approximately -1.2 eV and -2.1 eV below the conduction band minimum[21], primarily formed by OVs. Therefore, most electrons confined in deep levels can be excited to the conduction band by the light illumination. After switching off the light, photo-induced carriers are re-trapped by defect states, causing $R_S$ to increase. This recombination process, known as PPC[26], exhibits a relaxation trend similar to $V_G$-induced trapping but with significantly greater intensity and duration. This difference arises because the large relaxation induced by light involves the re-trapping of deep defect states, whereas $V_G$-induced trapping primarily affects shallow energy levels. To quantify this light-induced PPC, the $R_S$ recovery after the light illumination is also described by the same KWW function (Eq. 1), yielding a relaxation time constant $\tau_l = 9$ h and a stretching exponent $\beta_l = 0.2$. The extremely long $\tau_l$ underscores the dominance of deep traps in the PPC process, while the small value of $\beta$ indicates a broad energy distribution of these trapping centers. The pronounced contrast between the $V_G$-induced and light-induced relaxation parameters highlights the difference between deep and shallow defect states. Furthermore, as shown later, the $V_G$-induced $R_S$ variation evolves with the ongoing $R_S$ relaxation after light illumination, leading us to infer that $V_G$-induced electron trapping can be effectively modulated by the filling state of deep-level states.

B. The coupling between PPC and gating effect

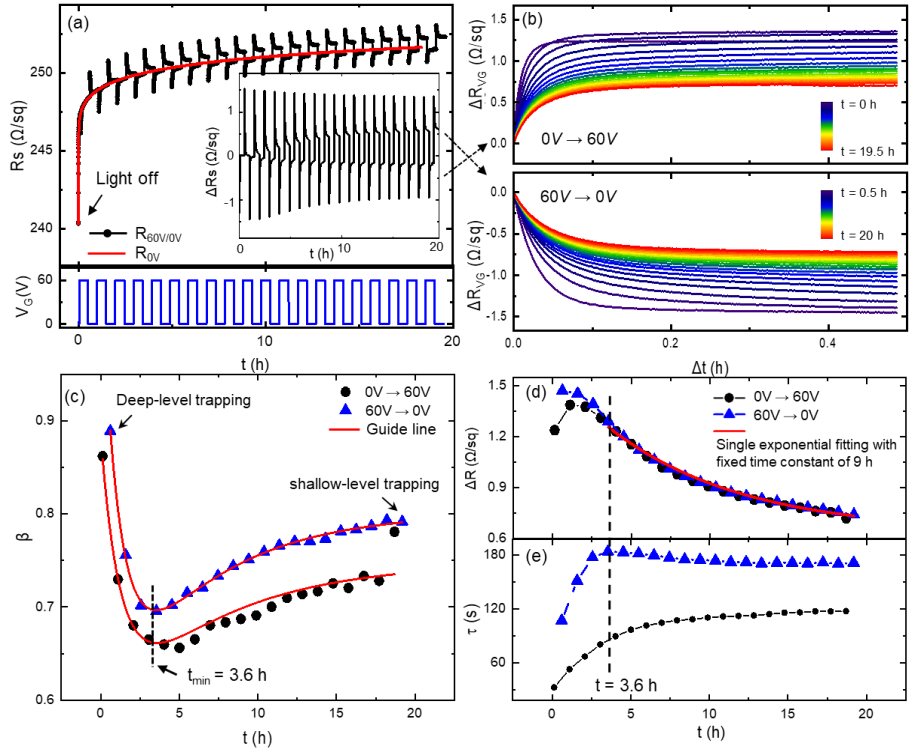

Fig. 3 (a) the $R_s$ variation with (black line) and without (red line) periodic $V_G$ steps after light is switched off. The fitted time constant $\tau_l$ of the red line is 9 hours. The lower panel is the corresponding gate voltage sweeping protocol. The inset shows the normalized result by subtracting black line and red line. (b) the $R_S$ variation extracted from the inset of (a), the upper and lower panel respectively represents the $\Delta R_{V_G}$ induced by positive (0→60 V) and negative (60→0 V) $V_G$ steps. All the $\Delta R_{V_G}$ can be well fitted by Eq. (1). (c-e) the time dependence of the fitting parameters ($\Delta R$, $\tau$, $\beta$). The red line in (d) is the fitting curve of a single exponential function with a fixed time constant of 9 h.

Figure 3(a) shows $R_S$ variation under periodic $V_G$ steps applied during the PPC relaxation (black line), the corresponding $V_G$ sweeping protocol is indicated by the blue line in the lower panel. The sequence consists of 20 $V_G$ steps and each with a duration of 0.5 hours. The red line represents the original $R_S$ relaxation of PPC in the absence of $V_G$. The application of periodic $V_G$ causes quasi-periodic steps in $R_S$, with the amplitude of the $V_G$-induced resistance relaxation ($\Delta R_{V_G}$) exhibiting a gradual temporal evolution. To clarify the time dependence of $\Delta R_{V_G}$, we subtracted the original PPC relaxation (red curve) from the $R_S$ measured under $V_G$ steps (black curve). The normalized result is shown in the inset of Fig. 3(a). The $V_G$-induced $R_S$ variation comprises two components: a transient response due to electrostatic charging, and a relaxation component arising from electron trapping and de-trapping. The transient component remains nearly constant throughout the post-illumination relaxation, whereas the trapping-related relaxation shows a pronounced time dependence. The relaxation component $\Delta R_{V_G}$ extracted at different times is summarized in Fig. 3(b). All relaxation profiles can be well fitted by the KWW function (Eq. 1), confirming that the underlying trapping mechanism persists during the PPC recovery.

The time dependence of the stretching exponent $\beta$ of Eq. (1) is shown in Fig. 3(c). The $\beta$ exhibit a non-monotonic variation over time, which indicates the gradual transition from deep- to shallow-level dominated trapping during the PPC recovery. As previously noted, $\beta$ reflects the width of the trap energy distribution, with values closer to 1 indicating a narrower distribution. At the initial seconds after light illumination ($t \approx 0$ h), a relatively high $\beta$ value suggests that the trapping process is dominated by the deep-level defects. As these deep states are progressively refilled by photo-excited carriers during PPC relaxation, $\beta$ decreases and reaches a minimum at $t_{min} = 3.6$ h. This minimum corresponds to the broadest distribution of active trap energies, marking a transitional regime where both deep and shallow states participate in the trapping process. Following an extended relaxation period, as the deep-level states become largely occupied and the interface approaches a quasi-equilibrium condition, the $V_G$-induced trapping is increasingly governed by shallower defect states, which is reflected in the subsequent recovery of $\beta$ toward higher values.

The evolution of the trapping energy distribution is further elucidated by the accompanying changes in amplitude $\Delta R$ and time constant $\tau$, as shown in Fig. 3(d) and (e). Both $\Delta R$ and $\tau$ exhibit an exponential decay over time, closely linked to the filling dynamics of deep-level states of PPC. Before the transitional point at $t = 3.6$ h, $\Delta R$ and $\tau$ show noticeable variation with time, likely resulting from the light-induced modifications of the OV-related trapping landscape. Meanwhile, the temporal evolution of the $\Delta R$ after $t = 3.6$ h can be well described by a single exponential function with a fixed time constant of 9 hours (red line in Figs. 3(d)), matching the PPC relaxation time constant. This concordance indicates that the $V_G$-induced trapping effect is modulated by the recombination kinetics of photo-induced carriers.

## C. Variation of electron trapping at different temperatures

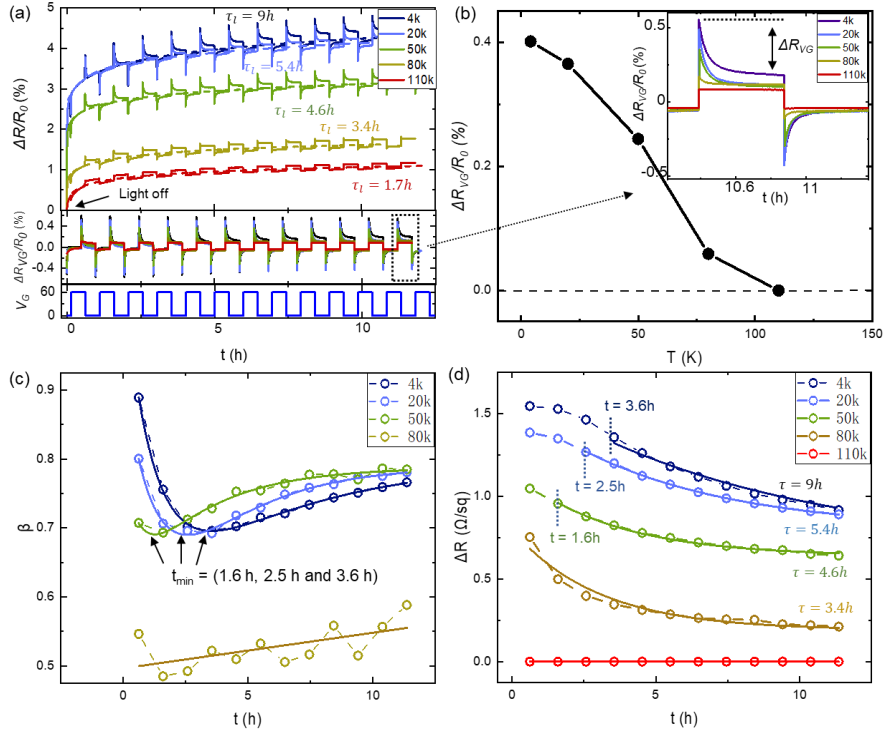

Fig .4 (a) Temperature dependence of $Rs$ with (solid line) and without (dashed line) periodic $V_G$ steps after light is switched off, all the dashed lines can be fitted by Eq. 2 and the fitted time constant $\tau_l$ is varying from 8.5 h at 4 K to 0.7 h at 300 K. the lower panel is the normalized result ($\Delta R_{VG}$) by subtracting solid line and dashed line. (b) the temperature dependence of the amplitude of $\Delta R_{VG}$. (c) The time dependence of $\beta$ at different temperatures, the minimum point at 4K, 20K, 50 K are respectively 1.6 h, 2.5 h and 3.6 h. (d) The time dependence of the $\Delta R_{VG}$ at different temperatures, the solid lines are the fitting curve of a single exponential function with the same time constants $\tau_l$ of (a), which varies from 9 h to 1.7 h.

To investigate the evolution of electron trapping and PPC across the ferroelastic transition of STO, we performed transport measurements at temperatures ranging from 4 K to 110 K. As shown in Fig. 4(a), both light- and $V_G$-induced $R_S$ relaxation amplitudes decrease with increasing temperature. This suppression can be attributed to two primary factors: firstly, elevated temperatures enhance the ionization of oxygen vacancies, thereby reducing the available population of empty traps and consequently the relaxation amplitude. Secondly, increased thermal activation energy alters the dominant re-trapping mechanism from slow quantum tunneling to a much faster thermal activation process, leading to a significant reduction in the observed relaxation time[27]. To isolate the $V_G$-induced $R_S$ variation, all $R_S$ data are subtracted by the baseline resistance without $V_G$ (dashed line in Fig. 4a), the normalized data are shown in the lower panel of Fig. 4(a). The trapping effect diminishes with increasing temperature and eventually vanishes at 110 K, which coincides with the ferroelastic phase transition temperature of STO. Below 105 K, STO undergoes an anti-ferrodistortive transition, forming a multi-domain state with ferroelastic twin walls[6]. These domain boundaries exhibit strain gradients and built-in polarization field[28], creating electrostatic potentials that attract both mobile carriers and oxygen vacancies[29, 30]. The twin walls act as "highways" for both defects and carriers, where OVs preferentially migrate and accumulate due to reduced migration barriers, forming OV clusters that function as efficient electron traps[31]. As previously discussed, the $V_G$-induced trapping process originates from OV clusters associated with ferroelastic twin walls. The density of twin walls decreases with increasing temperature and vanishes above the transition point[32]. Consequently, OV clustering is also suppressed, explaining the declining electron trapping observed in Fig. 4(b).

All the relaxation of normalized $\Delta R_{VG}$ at different temperature (middle panel in Fig. 4a) are well fitted by Eq. (1), the variation of fitting parameters ($\beta$ and $\Delta R$) is shown in Fig. 4(c) and (d). The stretching exponent $\beta$ remains relatively constant between 4 K and 50 K, indicating that the shallow defect states probed by $V_G$ maintain a similar energy distribution over this range and remain effective in trapping mobile carriers. With increasing temperature, the minimum point of $\beta$ ($t_{min}$) shifts to shorter times, reflecting an acceleration of the transition from deep- to shallow-level dominated trapping. Furthermore, the lower value of $\beta$ observed at 80 K suggests a broadening of the effective trap distribution, likely due to the thermal activation of additional shallow defect states that participate in the trapping process. Besides, the time dependence of $\Delta R$ show a similar decreasing trend at different temperature, and the part after the transitional point ($t > t_{min}$) can be well described by a single exponential function (solid lines in Fig. 4(c)). Also, the time constant for each exponential function is the same as the time constant $\tau_l$ derived from the light-induced $R_S$ relaxation (dashed line in Fig. 4a), which varies from 9 h (4 K) to 1.7 h (110 K). This consistent fitting further supports the conclusion that trapping induced by $V_G$ is modulated by the relaxation of PPC, i.e., the filling of deep-level states in the band gap.

### D. Schematic illustration of the trapping mechanism under PPC and $V_G$

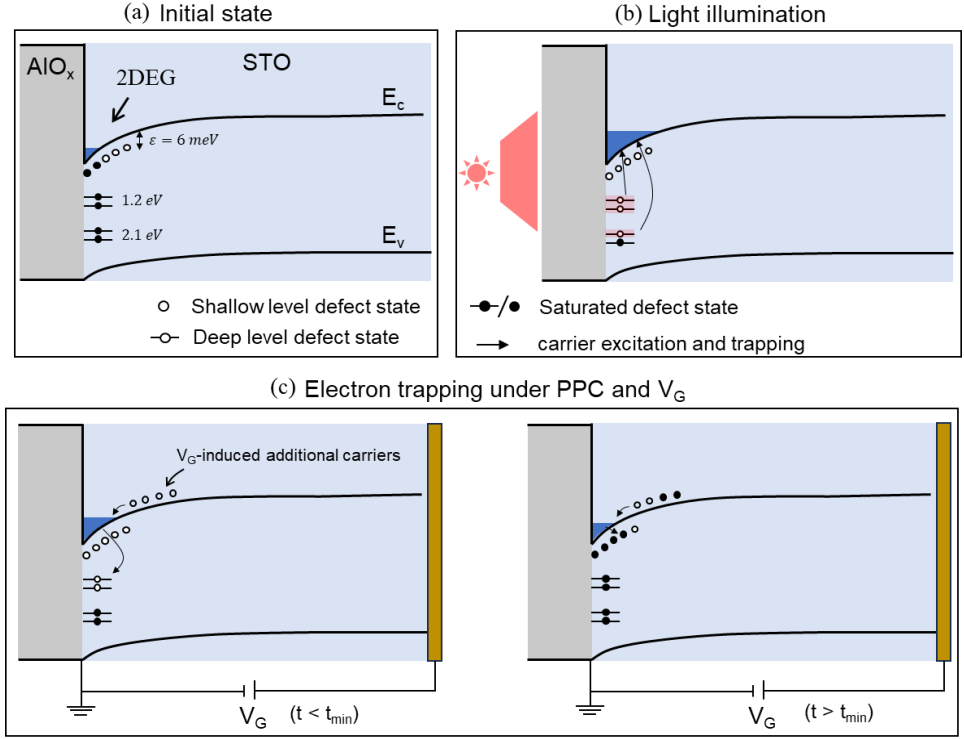

Fig. 5 (a) The initial state near the interface, where part of the shallow defect level are empty due to the thermal excitation. (b) Almost all the carriers within the band gap are excited to the conduction band under light illumination. (c) Before the transition point $t_{min}$, all the photo- and $V_G$-induced carriers are trapped by the deep-level states. After a long relaxation period (t ≫ $t_{min}$), all the deep-level traps are refilled. At this time, some of the $V_G$-induced carriers are quickly trapped by shallow-level defects, while the rest contribute to the 2DEG.

Based on the above experimental evidence, we propose a model involving multiple energy levels of interface trap states to explain the observed phenomena (Fig. 5). Below the STO ferroelastic transition temperature, three main defect types exist near the STO/AlO$_x$ interface: ferroelastic twin walls, isolated OVs, and OV clusters[23]. These defects constitute a spectrum of trap states with varying energy depths (from shallow to deep) and spatial distributions (Fig. 5a). Shallow-level traps, located near the conduction band bottom[20], have electrons easily excited by thermal energy or electric fields, leading to short trapping/de-trapping time constants (seconds to minutes). Deep-level traps, located farther from the conduction band[21], bind electrons firmly, requiring higher energy for release, and exhibit very long time constants (hours or days). Light illumination at 620 nm (~2.0 eV) excites electrons from deep levels into the conduction band, injecting them into the 2DEG (Fig. 5b). After the light is turned off, the system relaxes towards a new equilibrium. Free electrons are slowly re-trapped by the empty deep-level traps, a slow process requiring potential barrier crossing. This results in the observed slow $R_S$ relaxation (PPC), during which the system is in a non-equilibrium steady state with continuously changing deep-level trap filling. During this period, the $R_S$ disturbance induced by $V_G$ is transient, as any additional carriers introduced by $V_G$ are quickly trapped by the abundant empty deep-level traps, rendering the $V_G$ effect non-persistent (left panel of Fig. 5c). After a long relaxation period, all the deep-level traps are refilled. The interface trap system approaches thermal equilibrium. At this time, with most deep-level traps occupied, their capability for electron trapping is significantly reduced. When $V_G$ introduces additional electrons, some are quickly trapped by shallow-level defects, while the rest contribute to the 2DEG, resulting in a persistent $V_G$ effect (right panel of Fig. 5d).

## IV. CONCLUSIONS

In conclusion, we have systematically investigated the interaction between PPC and $V_G$-induced electron trapping at the STO/AlO$_x$ interface. Through time-dependent and temperature-dependent transport measurements, we unravel a unidirectional coupling where the PPC relaxation process acts as a master variable controlling the strength of $V_G$-induced trapping. The light-induced PPC reflects the slow filling of deep-level traps, while the transient response to $V_G$ reveals a faster charge redistribution process occurring against the background of the evolving deep-level trap population. During the PPC, the filling dynamics of deep-level traps dominate, whereas the $V_G$ primarily affects shallow defect energy levels. The interplay between light- and $V_G$-induced effects strongly depends on the system's history, specifically the current filling degree of the traps. Once deep-level traps are saturated, the control exerted by $V_G$ on shallow levels and band bending becomes clearly evident.

Our work demonstrates that a prior optical stimulus can set the state for subsequent electrical control. This concept of optically modulated electron trapping provides a new paradigm for controlling interface properties in complex oxides and offers a solid foundation for novel device applications.

This work was supported by NSFC under Grant No. 12274060.

Reference

(1) Gao, S. L.; Qiu, L. P.; Zhang, J.; Han, W. P.; Ramakrishna, S.; Long, Y. Z. Persistent Photoconductivity of Metal Oxide Semiconductors. *Acs Applied Electronic Materials* **2024**, *6* (3), 1542-1561. DOI: 10.1021/acsaelm.3c01549.

(2) Tarun, M. C.; Selim, F. A.; McCluskey, M. D. Persistent Photoconductivity in Strontium Titanate. *Physical Review Letters* **2013**, *111* (18). DOI: 10.1103/PhysRevLett.111.187403.

(3) Lei, Y.; Li, Y.; Chen, Y. Z.; Xie, Y. W.; Chen, Y. S.; Wang, S. H.; Wang, J.; Shen, B. G.; Pryds, N.; Hwang, H. Y.; Sun, J. R. Visible-light-enhanced gating effect at the LaAlO3/SrTiO3 interface. *Nature Communications* **2014**, *5*. DOI: 10.1038/ncomms6554.

(4) Jin, K. X.; Lin, W.; Luo, B. C.; Wu, T. Photoinduced modulation and relaxation characteristics in LaAlO3/SrTiO3 heterointerface. *Scientific Reports* **2015**, *5*. DOI: 10.1038/srep08778.

(5) Ohtomo, A.; Hwang, H. Y. A high-mobility electron gas at the LaAlO3/SrTiO3 heterointerface. *Nature* **2004**, *427* (6973), 423-426. DOI: 10.1038/nature02308.

(6) Pai, Y. Y.; Tylan-Tyler, A.; Irvin, P.; Levy, J. Physics of SrTiO3-based heterostructures and nanostructures: a review. *Reports on Progress in Physics* **2018**, *81* (3). DOI: 10.1088/1361-6633/aa892d.

(7) Tebano, A.; Fabbri, E.; Pergolesi, D.; Balestrino, G.; Traversa, E. Room-Temperature Giant Persistent Photoconductivity in SrTiO3/LaAlO3 Heterostructures. *Acs Nano* **2012**, *6* (2), 1278-1283. DOI: 10.1021/nn203991q.

(8) Rastogi, A.; Pulikkotil, J. J.; Budhani, R. C. Enhanced persistent photoconductivity in δ-doped LaAlO3/SrTiO3 heterostructures. *Physical Review B* **2014**, *89* (12). DOI: 10.1103/PhysRevB.89.125127.

(9) Rastogi, A.; Pulikkotil, J. J.; Auluck, S.; Hossain, Z.; Budhani, R. C. Photoconducting state and its perturbation by electrostatic fields in oxide-based two-dimensional electron gas. *Physical Review B* **2012**, *86* (7). DOI: 10.1103/PhysRevB.86.075127.

(10) Yin, C. H.; Smink, A. E. M.; Leermakers, I.; Tang, L. M. K.; Lebedev, N.; Zeitler, U.; van der Wiel, W. G.; Hilgenkamp, H.; Aarts, J. Electron Trapping Mechanism in LaAlO3/SrTiO3 Heterostructures. *Physical Review Letters* **2020**, *124* (1). DOI: 10.1103/PhysRevLett.124.017702.

(11) Hong, Y. P.; Jia, J. S.; Zhang, Z. T.; Wang, S.; Dou, R. F.; Nie, J. C.; Xiong, C. M. Electrostatic gating enhanced persistent photoconductivity at the LaAlO3/SrTiO3 interface. *Journal of Applied Physics* **2021**, *129* (11). DOI: 10.1063/5.0040891.

(12) Lu, H. L.; Liao, Z. M.; Zhang, L.; Yuan, W. T.; Wang, Y.; Ma, X. M.; Yu, D. P. Reversible insulator-metal transition of LaAlO3/SrTiO3 interface for nonvolatile memory. *Scientific Reports* **2013**, *3*. DOI: 10.1038/srep02870.

(13) Goyal, S.; Wadehra, N.; Chakraverty, S. Tuning the Electrical State of 2DEG at LaVO3-KTaO3Interface: Effect of Light and Electrostatic Gate. *Advanced Materials Interfaces* **2020**, *7* (16). DOI: 10.1002/admi.202000646.

(14) Rödel, T. C.; Fortuna, F.; Sengupta, S.; Frantzeskakis, E.; Le Fèvre, P.; Bertran, F.; Mercey, B.; Matzen, S.; Agnus, G.; Maroutian, T.; et al. Universal Fabrication of 2D Electron Systems in Functional Oxides. *Advanced Materials* **2016**, *28* (10), 1976-1980. DOI: 10.1002/adma.201505021.

(15) Vaz, D. C.; Noël, P.; Johansson, A.; Göbel, B.; Bruno, F. Y.; Singh, G.; McKeown-Walker, S.; Trier, F.; Vicente-Arche, L. M.; Sander, A.; et al. Mapping spin-charge conversion to the band structure in a topological oxide two-dimensional electron gas. *Nature Materials* **2019**, *18* (11), 1187-+. DOI: 10.1038/s41563-019-0467-4.

(16) Vaz, D. C.; Lesne, E.; Sander, A.; Naganuma, H.; Jacquet, E.; Santamaria, J.; Barthelemy, A.; Bibes, M. Tuning Up or Down the Critical Thickness in LaAlO3/SrTiO3 through In Situ Deposition of Metal Overlayers. *Advanced Materials* **2017**, *29* (28). DOI: 10.1002/adma.201700486.

(17) Vicente-Arche, L. M.; Mallik, S.; Cosset-Cheneau, M.; Noël, P.; Vaz, D. C.; Trier, F.; Gosavi, T. A.; Lin, C. C.; Nikonov, D. E.; Young, I. A.; et al. Metal/SrTiO3 two-dimensional electron gases for spin-to-charge conversion. *Physical Review Materials* **2021**, *5* (6). DOI: 10.1103/PhysRevMaterials.5.064005.

(18) Christensen, D. V.; Steegemans, T. S.; Pomar, T. D.; Chen, Y. Z.; Smith, A.; Strocov, V. N.; Kalisky, B.; Pryds, N. Extreme magnetoresistance at high-mobility oxide heterointerfaces with dynamic defect tunability. *Nature Communications* **2024**, *15* (1). DOI: 10.1038/s41467-024-48398-8.

(19) Liu, Z. Q.; Leusink, D. P.; Wang, X.; Lü, W. M.; Gopinadhan, K.; Annadi, A.; Zhao, Y. L.; Huang, X. H.; Zeng, S. W.; Huang, Z.; et al. Metal-Insulator Transition in SrTiO3-x Thin Films Induced by Frozen-Out Carriers. *Physical Review Letters* **2011**, *107* (14). DOI: 10.1103/PhysRevLett.107.146802.

(20) Zeng, S. W.; Yin, X. M.; Herng, T. S.; Han, K.; Huang, Z.; Zhang, L. C.; Li, C. J.; Zhou, W. X.; Wan, D. Y.; Yang, P.; et al. Oxygen Electromigration and Energy Band Reconstruction Induced by Electrolyte Field Effect at Oxide Interfaces. *Physical Review Letters* **2018**, *121* (14). DOI: 10.1103/PhysRevLett.121.146802.

(21) Schütz, P.; Christensen, D. V.; Borisov, V.; Pfaff, F.; Scheiderer, P.; Dudy, L.; Zapf, M.; Gabel, J.; Chen, Y. Z.; Pryds, N.; et al. Microscopic origin of the mobility enhancement at a spinel/perovskite oxide heterointerface revealed by photoemission spectroscopy. *Physical Review B* **2017**, *96* (16). DOI: 10.1103/PhysRevB.96.161409.

(22) Biscaras, J.; Hurand, S.; Feuillet-Palma, C.; Rastogi, A.; Budhani, R. C.; Reyren, N.; Lesne, E.; Lesueur, J.; Bergeal, N. Limit of the electrostatic doping in two-dimensional electron gases of LaXO3(X = Al, Ti)/SrTiO3. *Scientific Reports* **2014**, *4*. DOI: 10.1038/srep06788.

(23) Ojha, S. K.; Hazra, S.; Mandal, P.; Patel, R. K.; Nigam, S.; Kumar, S.; Middey, S. Electron Trapping and Detrapping in an Oxide Two-Dimensional Electron Gas: The Role of Ferroelastic Twin Walls. *Physical Review Applied* **2021**, *15* (5). DOI: 10.1103/PhysRevApplied.15.054008.

(24) Hüpkes, J.; Rau, U.; Kirchartz, T. Impact of Trap Depth on the Steady-State and Transient Photoluminescence in Halide Perovskite Films. *Advanced Energy Materials* **2025**. DOI: 10.1002/aenm.202503157.

(25) Guduru, V. K.; del Aguila, A. G.; Wenderich, S.; Kruize, M. K.; McCollam, A.; Christianen, P. C. M.; Zeitler, U.; Brinkman, A.; Rijnders, G.; Hilgenkamp, H.; Maan, J. C. Optically excited multi-band conduction in LaAlO3/SrTiO3 heterostructures. *Applied Physics Letters* **2013**, *102* (5). DOI: 10.1063/1.4790844.

(26) Di Gennaro, E.; di Uccio, U. S.; Aruta, C.; Cantoni, C.; Gadaleta, A.; Lupini, A. R.; Maccariello, D.; Marré, D.; Pallecchi, I.; Paparo, D.; et al. Persistent Photoconductivity in 2D Electron Gases at Different Oxide Interfaces. *Advanced Optical Materials* **2013**, *1* (11), 834-843. DOI: 10.1002/adom.201300150.

(27) Goyal, S.; Singh, A.; Tomar, R.; Kaur, R.; Bera, C.; Chakraverty, S. Persistent photoconductivity at LaVO 3?SrTiO 3 interface. *Solid State Communications* **2020**, *316*. DOI: 10.1016/j.ssc.2020.113930.

(28) Frenkel, Y.; Haham, N.; Shperber, Y.; Bell, C.; Xie, Y. W.; Chen, Z. Y.; Hikita, Y.; Hwang, H. Y.; Salje, E. K. H.; Kalisky, B. Imaging and tuning polarity at SrTiO3 domain walls. *Nature Materials* **2017**, *16* (12), 1203-+. DOI: 10.1038/nmat4966.

(29) Krantz, P. W.; Chandrasekhar. Observation of Zero-Field Transverse Resistance in AlOx/SrTiO3 Interface Devices. *Physical Review Letters* **2021**, *127* (3). DOI: 10.1103/PhysRevLett.127.036801.

(30) Rusevich, L. L.; Tyunina, M.; Kotomin, E. A.; Nepomniashchaia, N.; Dejneka, A. The electronic properties of SrTiO3-δ with oxygen vacancies or substitutions. *Scientific Reports* **2021**, *11* (1). DOI: 10.1038/s41598-021-02751-9.

(31) Lyzwa, F.; Pashkevich, Y. G.; Marsik, P.; Sirenko, A.; Chan, A.; Mallett, B. P. P.; Yazdi-Rizi, M.; Xu, B.; Vicente-Arche, L. M.; Vaz, D. C.; et al. Non-collinear and asymmetric polar moments at back-gated SrTiO3 interfaces. *Communications Physics* **2022**, *5* (1). DOI: 10.1038/s42005-022-00905-3.

(32) Ma, H. J. H.; Scharinger, S.; Zeng, S. W.; Kohlberger, D.; Lange, M.; Stöhr, A.; Wang, X. R.; Venkatesan, T.; Kleiner, R.; Scott, J. F.; et al. Local Electrical Imaging of Tetragonal Domains and Field-Induced Ferroelectric Twin Walls in Conducting SrTiO3. *Physical Review Letters* **2016**, *116* (25). DOI: 10.1103/PhysRevLett.116.257601.